\documentclass[prd,aps,nofootinbib,onecolumn,showpacs,10pt]{revtex4}
\usepackage{graphicx,epsfig,amssymb}
\usepackage[utf8x]{inputenc}

\usepackage{epsf}
\usepackage{graphicx}

\def   \lsth     {littlest Higgs}

\newcommand\iden{\leavevmode\hbox{\small1\normalsize\kern-.33em1}}
\def \nn {\nonumber}

\def\pbg{\frac{pb}{GeV}}

\begin{document}


\title{Double charged scalars of littlest higgs model in ee colliders\footnote{Talk given in "Third International Workshop on Prospects for Charged Higgs Discovery at Colliders - CHARGED2010,
		September 27-30, 2010, 
		Uppsala Sweden"}}



\author{Ay\c{s}e \c{C}a\~{g}ıl}
\affiliation{\vspace*{0.1in} Department of Physics, Middle East
Technical University,\\06531 Ankara, Turkey\\email:ayse.cagil@cern.ch}

\begin{abstract}

Little Higgs models contain heavy scalars in their content of particles as a result of extended symmetry group of SM. In the littlest
Higgs model of little Higgs models there exists a new heavy scalar triplet.The physical states of this triplet contains a double
charged scalar, a single charged scalar, as well as a neutral scalar and a neutral pseudo scalar. In little Higgs models a Majorana type
mass term can also be implemented in Yukawa Lagrangian, resulting lepton flavor violation.

In this talk the pair productions of double charged scalars in the context of littlest Higgs model in $e^+e^-$ colliders are presented. Also the
final signatures of double charged scalars will be analyzed depending on lepton flavor violation parameters. Finally it will be presented that if
there is lepton flavor violation, double charged scalars can be observed without any SM background in $e^+e^-$ colliders with a collider
signal of four leptons, otherwise if there is no lepton flavor violation they can be reconstructed with a background analysis.


\end{abstract}
\maketitle


\section{Introduction}
In litarature there are several models in which
triplet scalars occur , ranging from left-right symmetric models
to Little Higgs models. Complex scalar triplets with a hypercharge Y = 2 are
particularly rich in this context in terms of their phenomenological implications, especially in the aspect of lepton number violation.
    
In this talk I will present littlest Higgs model\cite{lh1}, one of the little Higgs models\cite{lhmodels1}, having a complex SU(2) scalar sector, and production of heavy double charged scalars via $e^+e^-\to \phi^{++}\phi^{--}$ and $e^+e^-\to Z_L \phi^{++}\phi^{--}$  channels in $e^+e^-$ colliders. This talk is a summary of the double charged productions discussed in previous works\cite{cagil1,cagil2}. Reviews on little Higgs models can be found in numerous works\cite{reviews}.

\section{Littlest Higgs Model}  
The main motivation of Little Higgs models are to overcome hierarchy problem. To do so these models use collective symmetry breaking mechanism that softens radiative corrections contributing to Higgs mass. In the littlest Higgs model, one of the minimal Little Higgs model,   global
symmetry $SU(5)$ is broken spontaneously to $SO(5)$ at an energy
scale $f\sim 1TeV$ leaving $14$ Nambu Goldstone bosons(NGB). Global  $SU(5)$ in the model contains a weakly gauged subgroup 
$[SU(2)_1\otimes U(1)_1]\otimes[SU(2)_2\otimes U(1)_2]$ , thus eight gauge bosons. As the global symmetry $SU(5)$ is broken spontaneously to $SO(5)$ triggered by a chosen vacuum basis, gauged subgroup 
$[SU(2)_1\otimes U(1)_1]\otimes[SU(2)_2\otimes U(1)_2]$ is broken to its diagonal subgroup 
 $(SU(2)\otimes U(1))$ of SM. As a result gauge bosons get mixings and some gain mass by eating four of the NGBs. The remaining NGBs at this stage are  a scalar triplet $\phi$ and a scalar doublet  $h$. The usual electroweak symmetry breaking occurs by the vacuum expectation value(VEV) of the
Higgs potential written by Coleman Weinberg method for scalars. 
With EWSB vector bosons get extra mixings due to vacuum expectation
values of $h$ doublet and $\phi$ triplet; $v$ and $v'$ respectively.

The particle content after EWSB contains SM vector bosons ; $Z_L$, $W_L$, $A_L$ , and extra massive vector bosons ; $Z_H$, $W_H$, $A_H$ . Diagonalizing the mass matrix for
scalars the physical states are found to be the SM Higgs scalar $H$,
the neutral scalar $\phi^0$, the neutral pseudo scalar $\phi^P$, and
the charged scalars $\phi^+$ and $\phi^{++}$. The masses of these
scalars are degenerate, and in terms of Higgs mass can be expressed
as:
\begin{eqnarray}
    M_\phi =\frac{\sqrt{2} f}{v\sqrt{1-(\frac{4 v'
    f}{v^2})^2}}M_H,
\end{eqnarray}
In the model new vector bosons cancel out the divergences coming to Higgs mass from SM vector bosons loops, new scalars cancel out the divergences coming to Higgs mass from Higgs self loop. To cancel out the quadratic divergence coming to Higgs mass from top quark loop, a new extra singlet the $T$ quark is introduced in the model by hand.

The scalar fermion interactions in the model are written in Yukawa
Lagrangian preserving gauge symmetries of the model for SM leptons
and quarks, including the third generation having an extra singlet,
the $T$ quark. In this work, leptons are charged under both $U(1)$
groups, with corresponding hypercharges of $Y_1$ and $Y_2$. The
restriction for $Y_1$ and $Y_2$ is that $Y_1 + Y_2$ should reproduce
$U(1)_Y$ hypercharge ${\cal Y}$ of SM, thus $Y_1=x {\cal Y}$ and $Y_2=(1-x){\cal Y}$ can
be written. Due to gauge invariance, $x$ can be taken as
$3/5$\cite{B2csaki}. 

In littlest Higgs model, the symmetry braking scale $f$ and mixing angles between $U(1)$
groups and  $SU(2)$ groups; $s$ and $s'$ respectively, are free parameters and they are constrained by observables\cite{constraints}. 
The recent data from Tevatron and LEPII constrain the mass of the lightest heavy scalar as $M_{A_H}\gtrsim900GeV$\cite{TEVATRON}. In the original formulation of the littlest Higgs model, these data imposes strong constraints on symmetry breaking scale($f>3.5 - 4 TeV$). But in this work by gauging fermions in both $U(1)$ subgroups, fermion boson couplings are modified as done in\cite{B2csaki}. 
With this modification the symmetry breaking scale can be lowered to $f=1TeV$, which allows the mass of the $A_H$ to be at the order of few $GeV$s. In this case the allowed parameter region of the littlest Higgs 
model is  as follows. For low values of the symmetry breaking scale $1TeV
\leq f \leq 2 TeV$, mixing angles $s$ and $s'$ between gauge bosons
are in the range $0.8\leq s \leq 1$ and $0.6\leq s' \leq 0.7 $, and
for $ 2TeV \leq f \leq 3TeV$ they have acceptable values in the
range $0.65\leq s \leq 1$ and $0.4\leq s' \leq 0.9$. For the higher
values of the symmetry breaking scale $f\geq 3TeV$, the mixing
angles are less constrained, since the corrections to SM observables
from {\lsth} model comes in the form $\frac{v^2}{f^2}$ and higher
orders.

In the model for light fermions, a Majorano
type mass term can be implemented in Yukawa
Lagrangian\cite{thanlept1} which
results in lepton flavor violation by unit two, such as:
\begin{equation}\label{lepviol1}
    {\cal L}_{LFV} = iY_{ij} L_i^T \ \phi \, C^{-1} L_j + {\rm h.c.},
\end{equation}
where $L_i$ are the lepton doublets $\left(
                                       \begin{array}{cc}
                                         l &\nu_l \\
                                       \end{array}
                                     \right)$,
and $Y_{ij}$ is the Yukawa coupling with $Y_{ii}=Y$ and $Y_{ij(i\neq
j)}=Y'$ . The values of Yukawa couplings $Y$ and $Y'$ are restricted
by the current constraints on the neutrino
masses\cite{neutrinomass}, given as; $M_{ij}=Y_{ij}v'\simeq
10^{-10}GeV$\cite{thanlept1}. Since the vacuum expectation value
$v'$ has only an upper bound; $v'<1GeV$, $Y_{ij}$ can be taken up to
order of unity without making $v'$ unnaturally small.

The decay modes of littlest Higgs model scalars, including the lepton flavor violating decays, are studied in T.Han et al\cite{thanlept1}. Since these new scalars have lepton flavor
violating modes, their total widths will depend on the Yukawa
couplings $Y_{ii}=Y$ and $Y_{ij(i\neq j)}=Y'$.Depending on these mentioned parameters double charged scalars can decay into non leptonic particles WW pairs or leptons of same kind or leptons of different kind violating lepton number by 2. 

The decay width of
$\phi^{++}$ is given as\cite{thanlept1}:
\begin{eqnarray}\label{dwp2}
 \nn   \Gamma_{\phi^{++}}
    &\approx&  \frac{v^{\prime 2} M_{\phi}^3}{2 \pi v^4}+\frac{3}{8\pi } |Y|^2 M_\phi+\frac{3}{4\pi } |Y'|^2 M_\phi.
\end{eqnarray}

In the calculations, we ignored $v^2/f^2$ terms in
the couplings, since we are not dealing with the corrections to a SM
process. Further information on theory and calculations, and Feynman diagrams contributing the processes $e^{+}e^{-}\rightarrow Z_L \phi^{++} \phi^{--}$ and $e^{+}e^{-}\rightarrow \phi^{++} \phi^{--}$ can be found in \cite{cagil1,cagil2}.

\section{Results and Discussions}

\subsection{$e^{+}e^{-}\rightarrow Z_L \phi^{++} \phi^{--}$}

\begin{figure}[h]
\begin{center}
\includegraphics[width=6.5cm]{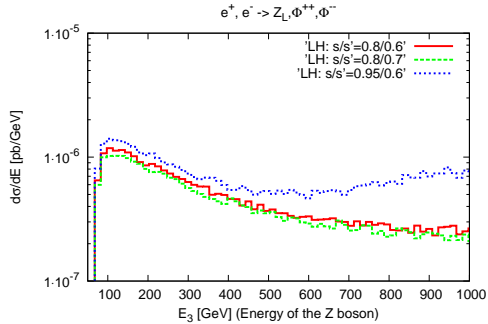}
\caption{Differential cross section vs. $E_Z$ graphs for process
$e^{+}e^{-}\rightarrow Z_L \phi^{++} \phi^{--}$ for the fixed
value of parameters $s/s':0.8/0.6,0.8/0.7,0.95/0.6$ at
$f=1000GeV$ at $\sqrt{s}=3TeV$.} \label{zppC1a}
\end{center}
\end{figure}

\begin{figure}[h]
\begin{center}
\includegraphics[width=7cm]{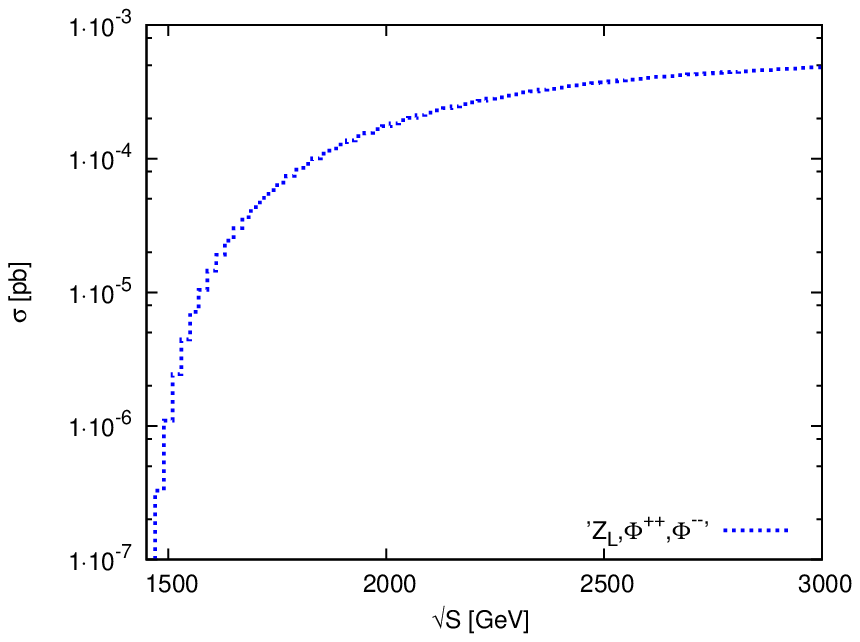}
\caption{Total cross section vs.
$\sqrt{S}$ graphs for the process $e^{+}e^{-}\rightarrow Z_L\phi^{++} \phi^{--}$ for some values of 
mixing angle parameters $s/s'$ when $f=1000GeV$.} \label{ZPPc2}
\end{center}
\end{figure}

For the associated pair production of doubly charged scalars within
$Z_L$ the differential cross sections with respect to $E_Z$ are
plotted in figure \ref{zppC1a} for different fixed values of
mixing angles at $\sqrt{s}=3TeV$, at symmetry breaking scale
$f=1TeV$. The dependence of total cross section on the center of
mass energy of this production process is plotted in figure
\ref{ZPPc2}, and the numerical values of total cross sections are
presented in table \ref{cseccharged} for parameters of interest.

\begin{table}[htb]
\begin{center}
\begin{tabular}{|c||c|c|}
  \hline $s/s'$  &$\sigma_{\phi^{++}\phi^{--}}$  & $\sigma_{Z_L\phi^{++}\phi^{--}}$ \\
  \hline
  \hline  $0.8/0.6$& $ 2.2\times 10^{-3}$  & $ 0.48\times 10^{-3}$ \\
  \hline  $0.8/0.7$&  $ 2.1\times 10^{-3}$&  $0.44\times 10^{-3}$ \\
  \hline  $0.95/0.6$& $ 2.7\times 10^{-3}$  &$ 0.78\times 10^{-3}$  \\
  \hline
\end{tabular}
\caption{The total cross sections in $pb$ for pair production of
charged scalars for $f=1TeV$ and at
$\sqrt{s}=3TeV$.} \label{cseccharged}
\end{center}
\end{table}

For the electroweak allowed parameters
$s/s'=0.8/0.6,0.8/0.7,0.95/0.6$ at $f=1TeV$ at $\sqrt{s}=3TeV$, the
differential cross section gets low values at the order of
$10^{-7}\pbg$. The resulting cross sections are calculated by
integrating over $E_Z$, and found about $0.4\sim0.8\times 10^{-3}pb$ (table
\ref{cseccharged}) resulting $40\sim80$ events per year for
integrated luminosity of $100fb^{-1}$. For $\sqrt{s}<2TeV$, this
production channel is not reachable. In the {\lsth} model,
$\phi^{++}$ has decays to charged vectors $W^+_L W^+_L$ and also to
leptons $l^+_i l^+_j$ proportional to squares of the values of the
Yukawa couplings; $|Y^2|$ for same families and $|Y'^2|$ for
different lepton families when lepton violating modes are
considered. So this channel provides final signals for doubly
charged scalar discovery and lepton flavor violation. The final
states of the doubly charged scalar pairs dominantly contain
leptonic modes $l_i l_i l^+_i l^+_j $, semi leptonic modes $l_i l_i
W^+_L W^+_L$ and to standard model charged vector pair $W^+_L W^+_L
W^-_LW^-_L$ depending on $Y$ and $Y'$, while $Z_L$ dominantly decays
to jets carrying the energy at the order of masses of the scalar
pair. For $f=1TeV$ the leptonic branching ratio of doubly charged
scalars can reach values close to $1$ for $Y\rightarrow 1$,
independent from $Y'$. If the value of the Yukawa coupling $Y$ is
high enough $(Y\sim 1)$, the number of final state lepton flavor
violating signals such as; $Z_L l_il_i l^+_j l^+_j$, can reach up to
$50$ events per year for luminosities of $100fb^{-1}$, which can be
directly detectable free from backgrounds.

\subsection{$e^{+}e^{-}\rightarrow \phi^{++} \phi^{--}$}

\begin{figure}[tbh]
\begin{center}
\includegraphics[width=7cm]{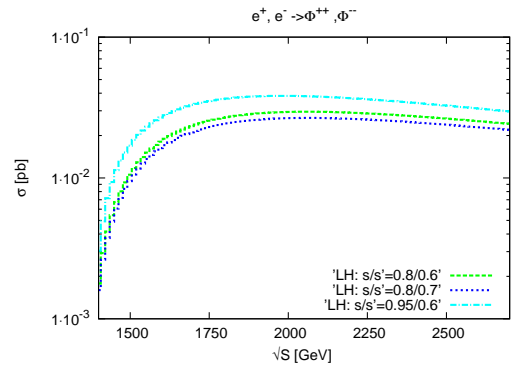}
\caption{Total cross section vs.
$\sqrt{S}$ graphs for the process $e^{+}e^{-}\rightarrow \phi^{++} \phi^{--}$ for some values of 
mixing angle parameters $s/s'$ when $f=1000GeV$.} \label{HPc2}
\end{center}
\end{figure}

For the process $e^{+}e^{-}\rightarrow \phi^{++} \phi^{--}$ the total cross section of the production event is examined 
for $f=1TeV$. The dependence of the total cross section of the
$e^{+}e^{-}\rightarrow \phi^{++} \phi^{--}$ process on $\sqrt{s}$ at
fixed values of the littlest Higgs model parameters are presented in
Fig. \ref{HPc2}. For the mixing angles $s/s'=0.8/0.6$, the production cross section is maximum 
with a value of $3\times10^{-2}pb$ for $f=1TeV$ at $\sqrt{S}>1.7TeV$. 
For the parameter sets $s/s':0.8/0.7,0.95/0.6$ for $f=1TeV$, the total cross section is is slightly lower but still 
in the order of 
$10^{-2}pb$ at $\sqrt{S}>1.7TeV$. Thus for an $e^+e^-$ collider with an integrated luminosity of $100fb^{-1}$, yearly $3000$ double charged pair production can be observed. 

\begin{figure}[hb]
\begin{center}
\includegraphics[width=8cm]{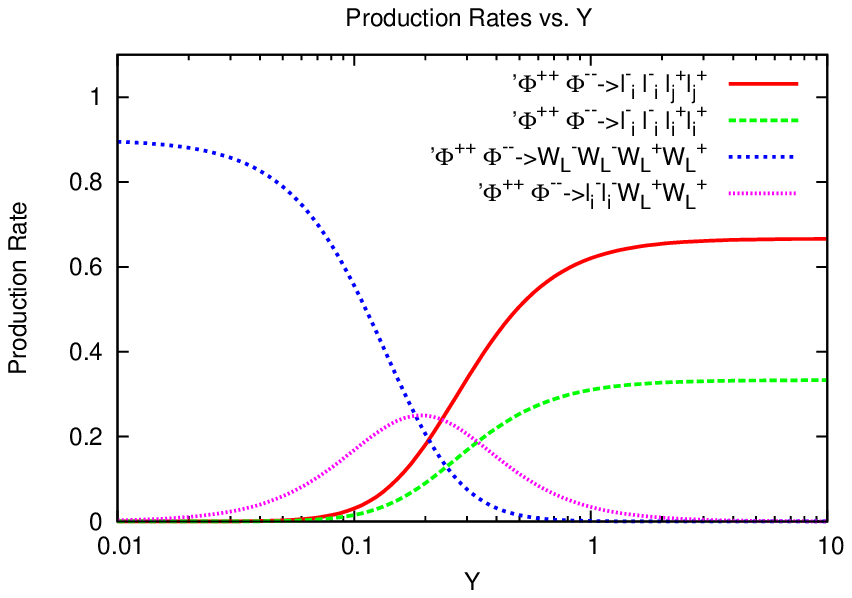}
 \vskip2mm \caption{Dependence of the final collider signatures of the process $e^+e^- \to \phi^{++}\phi^{--}$ on $Y$.} \label{BR2}
\end{center}
\end{figure}

It is seen from the final decay modes of the doubly charged pair given in Eq.\ref{dwp2} that the final state 
analysis is strongly dependent on the value of the Yukawa coupling $Y$, thus on the value of the triplet 
VEV $v'$. The dependence of the final collider signatures of the process $e^+e^- \to \phi^{++}\phi^{--}$ on $Y$ 
are plotted in figure \ref{BR2}.

For $Y<0.01(v'>10^{-8}GeV)$, the final decays of the doubly charged scalars are dominated by SM charged bosons, and the final collider signature will be $W^+_L W^+_L W^-_L W^-_L$. In this case with a subtraction from background, doubly charged scalars can be identified by reconstructing the same sign boson pairs.

For $0.01<Y<1(10^{-10}GeV<v'<10^{-8}GeV)$, the semi leptonic decay modes, $l_i l_i W_L^+ W_L^+$ will be observed. 
For $Y\sim 0.2$, the production rates for these modes are calculated as; $PR(Y\sim0.2)=0.2$, leading to $600$ collider signals 
per year at collider with a luminosity of $100fb^{-1}$ at $\sqrt{S}>1.7TeV$. The cleanest signal in this case will be observed when both $W_L^+$ decay into jets ($48\%$). In this scenario there will be yearly $280$ signal of two same sign leptons of same family plus jets, violating lepton number and flavor by two, which can be directly detectable free from any backgrounds.

The most interesting scenario happens when the Yukawa coupling is close to unity, $Y\sim 1(v'\sim 10^{-10}GeV)$. In this case all final doubly charged 
scalars will decay into same family leptons, and the final collider signal will be either $l_i l_i \bar{l_j} \bar{l_j}$  or 
$l_i l_i \bar{l_i} \bar{l_i}$. The mixed states $l_i l_j \bar{l_i} \bar{l_j}$ resulting from the final decays of 
the doubly charged scalars into different families of leptons are suppressed because of the low production rates due to the value of the Yukawa mixing coupling $Y'$.

The branching ratio into the final signal $l_i l_i \bar{l_j} \bar{l_j}$ when $Y\sim1$ is calculated as $PR(Y\sim1)=0.66$, which will 
give $1800$ observable signals per year at an integrated luminosity of $100fb^{-1}$ when $\sqrt{S}>1.7TeV$. 
For this final signal, the lepton flavor is violated explicitly, free from any backgrounds.  The double charged scalars in this case 
can be reconstructed from invariant mass distributions of same charged leptons.

For the final state $l_i l_i \bar{l_i} \bar{l_i}$, there will be additional observable $1000$ events per year. For this case lepton flavor violation can not be observed 
directly even if it happens via the lepton flavor violating decays of doubly charged scalars. Although this signal has a huge SM background, with a proper background analysis, 
the existence of doubly charged scalars and so on their lepton flavor violating decays can be identified.

\end{document}